\documentclass{iau} 
\usepackage{graphicx}
\usepackage{gensymb}
\usepackage{xmpmulti}
\usepackage{subfigure}
\usepackage{hyperref}

\title[Analysis of Magnetic Field Variations Produced by EEJ] 
{Analysis of  Magnetic Field Variations  Produced by Equatorial Electro-Jets.}
\author[Ericson L\'opez, Franklin Ald\'as \& Akimasa  Yoshikawa]   
{Ericson L\'opez$^1$
Franklin Ald\'as$^2$
 \and Akimasa  Yoshikawa $^3$}

\affiliation{$^1$Observatorio Astron\'omico de Quito, Escuela Polit\'ecnica Nacional, Quito, Ecuador \\ email: {\tt ericsson.lopez@epn.edu.ec} \\[\affilskip]
$^2$Observatorio Astron\'omico de Quito, Escuela Polit\'ecnica Nacional, Quito, Ecuador \\ email: {\tt franklin.aldas@epn.edu.ec} \\[\affilskip]
$^3$ International Center for Space Weather Science and Education, Kyushu University, Japan \\email: {\tt yoshi@geo.kyushu-u.ac.jp}}
\pubyear{2017}
\volume{xxx}  
\jname{ Space Weather of the Heliosphere: Processes and Forecasts}
\editors{Claire Foullon, \& Olga Malandraki}
\begin{document}

\maketitle

\begin{abstract}
The Equatorial Electrojet (EEJ) is a narrow band of electrons flowing from east to west at daytime at low latitudes. The electron current produces a magnetic field variation that can be measured at different latitudes. In this work, we have used the data analysis in order to quantify the solar and lunar contributions to those variations and study the morphology of the EEJ current. 
\keywords{Equatorial Electrojet, magnetic field, data analysis.}
\end{abstract}
\firstsection 
\section{Introduction}
Observations of the magnetic field at Huancayo in 1922 showed the existence of Equatorial Electrojet (EEJ), which is a narrow band of electrons traveling from east to west, 180 km above the surface of the Earth (\cite[Chapman, 1951]{Chapman}). 
Wind-driven currents, through the dynamo mechanism, produce an accumulation of positive charges at sunrise and negative at sunset (\cite[Huang, 2005]{Huang}), which results in considerable variations of the geomagnetic field. 
The variations of the magnetic field are no constant, instead, they depend on solar activity, geographical location and vary temporally with season (\cite[Rigoti, 1999]{Rigoti}).
\section{Data Set}
Ground magnetometers allow  a  deeper understanding of the terrestrial ionosphere. The Quito Astronomical Observatory, in collaboration with  the University of Kyushu,  has installed a Magnetic Acquisition System (MAGDAS) in Jerusalem, located 35 km northward from Quito, Ecuador. The magnetometer location has geographic coordinates GLAT: 0\degree 0' 8.67'' S, GLON: 78\degree 21'24.87'' W (\cite[L\'opez, 2014]{Lopez}), and it is operative from October 2012 until now. \\
We have used the horizontal component of the magnetic field (H) recorded by the Ecuadorian MAGDAS magnetometer in combination with the data downloaded  for free from the project: A Global Ground-Based Magnetometer Initiative, available at web page \\(\href{http://supermag.jhuapl.edu/mag/?}{http://supermag.jhuapl.edu/mag/?}),  during the period  since October 2012 to December 2016. The geographic and geomagnetic locations of the stations used in this work are indicated in the Table 1.
\begin{table}[h!b!t!]
\centering
\begin{tabular}{|l|c|c|c|c|c|}
\hline
  Station & Acronym & GLON &GLAT&MLON& MLAT   \\
  \hline
Vulcan &T03&247.02&50.37&-50.40&57.60\\  
Boulder&BOU&254.77&	40.13	&-38.68	&48.51\\
Fredericksburg&FRD&282.63	&38.20&	-0.64	&48.05\\
Fresno&FRN&240.30	&37.10	&-54.87&	42.65\\
Bay Sait Louis&BSL&270.37	&30.35&	-17.88&	40.69\\
Tucson&TUC&249.27&	32.17	&-43.96	&39.32\\
San Antonio&M08&261.39	&29.44	&-29.35	&38.74\\
San Juan&SJG&293.85	&18.11&	11.85	&26.75\\
Jerusalem &JRS&282.64&-0.0024&-5.9800&9.67\\
Kourou&KOU&307.27	&5.21&	24.46	&8.62\\
Huancayo&HUA&284.67&-12.05&-2.75&1.17\\
Villa Remedios&VRE&292.38&	-17.28&	3.97&	-4.53\\
Pilar&PIL&294.47&	-31.40&	4.76&	-18.39\\
Osorno&OSO&286.91	&-40.34	&-0.03	&-26.48\\
Punta Arenas&PAC&289.91&	-40.34&	1.97&	-26.64\\
Port Stanley &PST&302.11&-51.70&-38.96&-38.96\\
\hline
\end{tabular}
\caption{Geographic and geomagnetic positions of ground stations.}
\label{locaciones}
\end{table}
\section{Data Analysis}
The magnetic field measurements provided by the ground stations is the sum of multiple contributions, among the more significant are: the Earth main magnetic field, the secular current ($S_q$), the field aligned currents and so on. In order to study the cyclic variations, we have removed the non-periodic variations by correction of the  disturbance field ($\delta_H$) described by \cite[Yamazaki (2017)]{Yamazaki}. The zero level of secular variations ($S_q$) has been obtained using the hourly $D_{st}$ index  which provides a quantitative measure of geomagnetic disturbance, that can be correlated with other solar and geophysical parameters. For an station with a geomagnetic latitude $\lambda$, the disturbance field $\delta_{H}$ over H can be related by the expression $\delta_H=D_{st}*cos(\lambda)$ \cite[Takeda, 2002]{Takeda}.\\
{\underline{\it Fourier Series Analysis}}. Given the periodicity of the secular variations $ S_q $ of the terrestrial magnetic field, it is possible to expand it in Fourier series, it means to express $S_q$ as an infinite sum of sines and cosines. In general, in order to capture the solar variability, it is enough if we use the first four solar harmonics (i.e., 24, 12, 8 and 6 hours).\\
The mean values of average secular variations ($S_q$) as function of magnetic latitude (MLAT) are plotted in the Figure 1, where we can see that the EEJ current band is symmetric,  centered at $2.2\degree $  MLAT and with full-width at half-maximum of $ 18.12 \degree$.  
\begin{figure}[h!b!t!]
\begin{center}
\includegraphics[height=4.5cm, trim={0cm 0cm 0cm 0cm},clip]{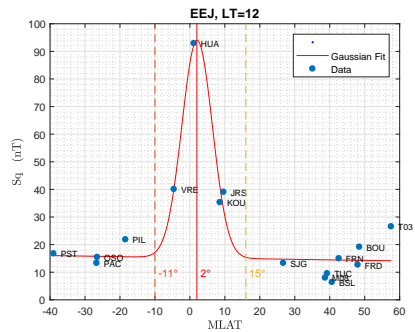}
\caption{Profile of secular variations ($S_q$) as function of the magnetic latitude (MLAT).}
\end{center}
\label{NEZ}
\end{figure}
{\underline{\it Lunar Contributions to Magnetic Field Variations}}. 
We define the lunar age $ \nu $ by analogy with the solar day. A lunar cycle ($ 29.5306$ solar days) has a duration of 24 lunar hours. This means that $0$ lunar hours represents the new moon and $12$ lunar hours represents the full moon. The Moon has a significant contribution during EEJ production, in order to quantify the lunar  contribution we have subtracted the first harmonic calculated through the Fourier series expansion from the original data.  The residual  can be  considered as the contribution of the Moon to the variation of the horizontal magnetic field. Then, the residuals are grouped by lunar age and local time (LT) as we show in the Figure 2. 
\begin{figure}[!h!b!t]
\begin{center}
\begin{tabular}{c c}
\resizebox{.36\linewidth}{!}{\subfigure{\includegraphics[trim={0cm 0cm 0cm 0cm},clip]{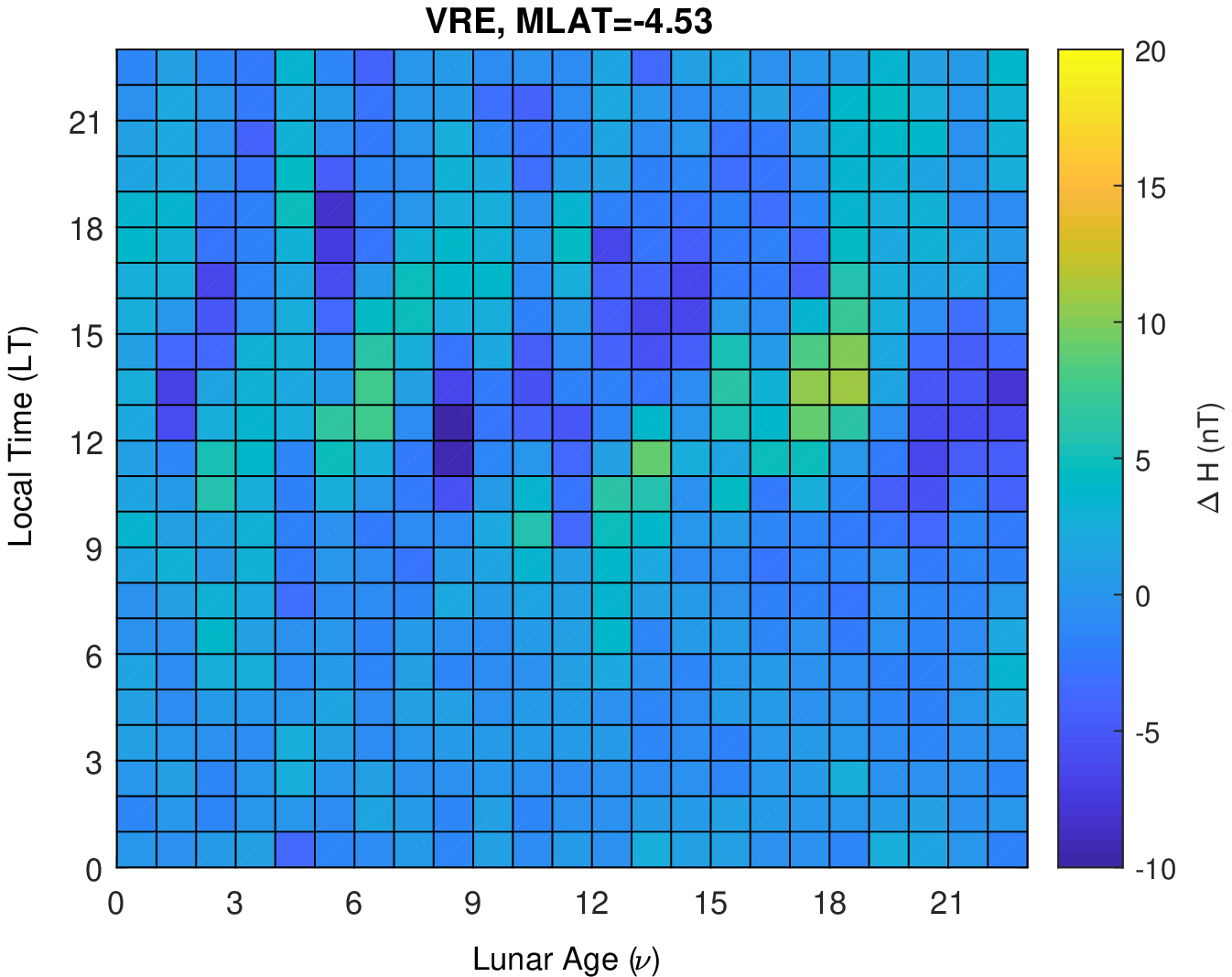}
                \label{fig:subfig1}}}&
\resizebox{.36\linewidth}{!}{\subfigure{\includegraphics[trim={0cm 0cm 0cm 0cm},clip]{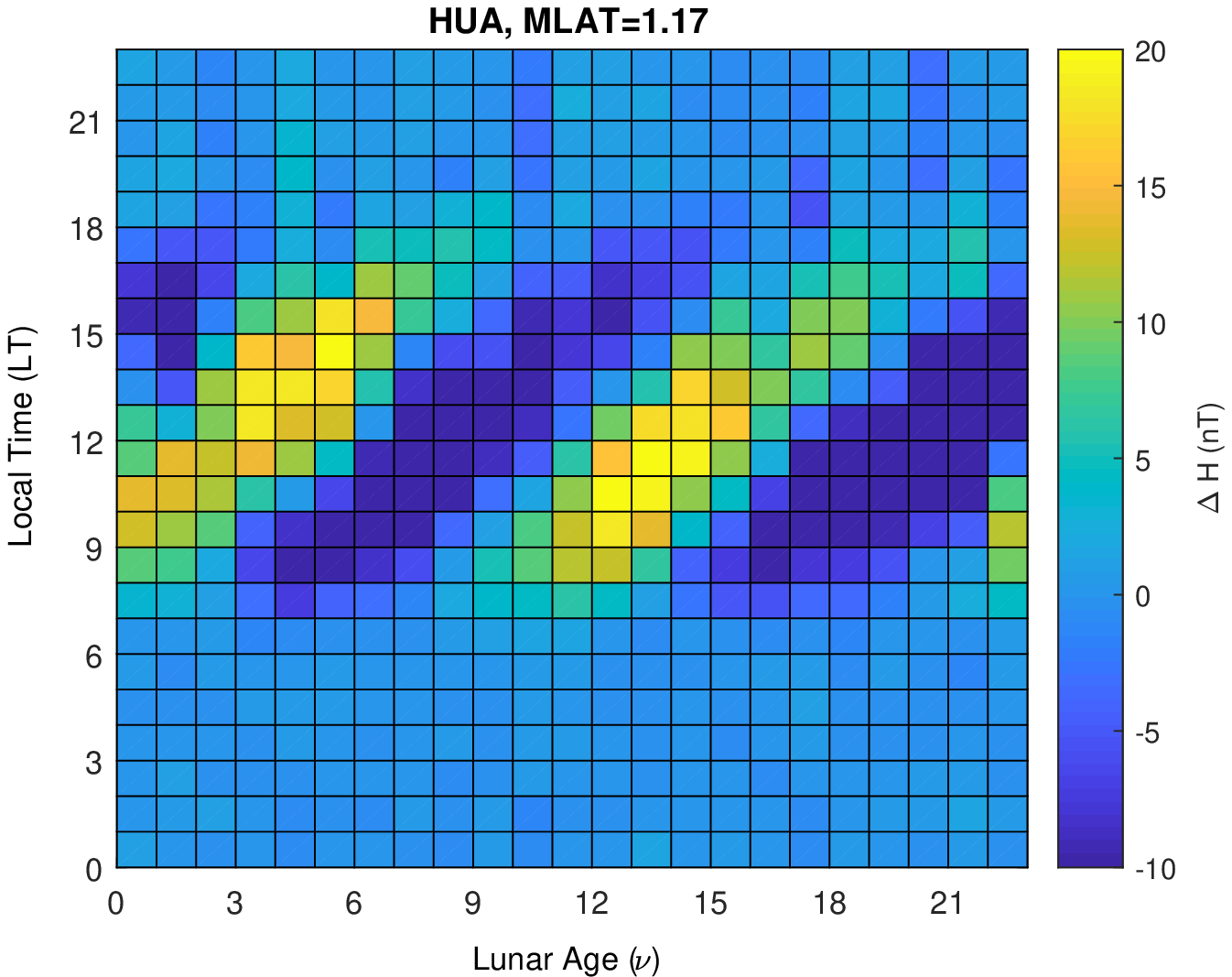}
                \label{fig:subfig1}}}\\
\resizebox{.36\linewidth}{!}{\subfigure{\includegraphics[trim={0cm 0cm 0cm 0cm},clip]{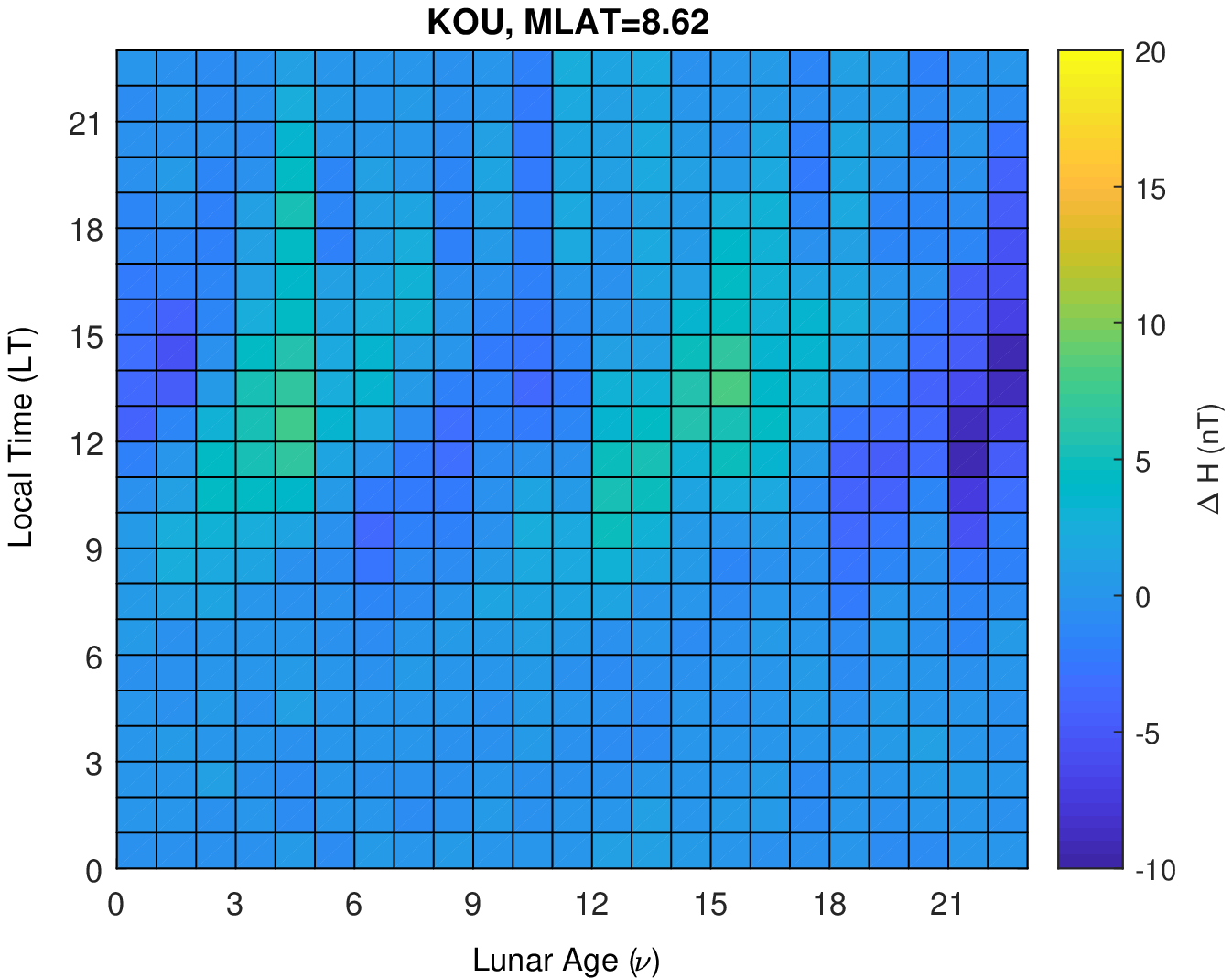}
                \label{fig:subfig1}}}&
\resizebox{.36\linewidth}{!}{\subfigure{\includegraphics[trim={0cm 0cm 0cm 0cm},clip]{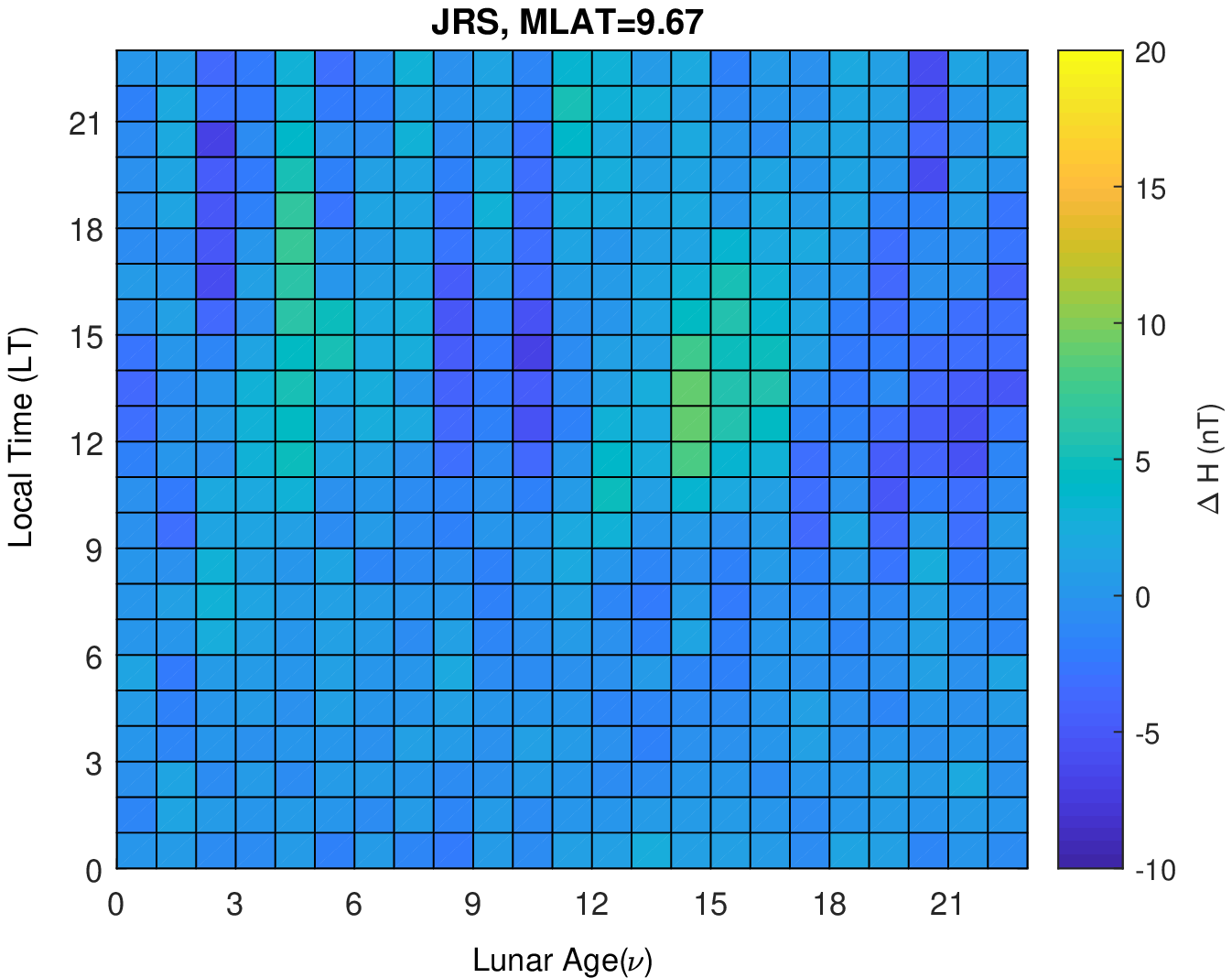}
                \label{fig:subfig1}}}\\
\end{tabular}
\caption{Lunar variations for four different stations located near to dip equator.}
\label{variablidad}
\end{center}
\end{figure}

\section{Conclusions}
From the Gaussian fit of the secular variations of the horizontal magnetic field component as a function of the geomagnetic latitude,  we are able to derive that the full-width at half-maximum for  average EEJ  on America's equatorial region is  $18.12 \degree$ . The mean EEJ current during the years 2012 - 2016 extends from $-11 \degree $ to $15 \degree$ magnetic latitudes and it is centered at $MLAT=2 \degree$. \\
We found a dependence on the lunar age and the day time in the electrojet generation. Our results are similar to those presented by Yamazaki  (\cite[Yamazaki 2017]{Yamazaki}), but derived by another method. Although, the Moon contribution is important in the EEJ formation, its contribution to the magnetic field is, in average, five times less important than the Sun contribution.


\begin{thebibliography}{}
\bibitem[Chapman (1951)]{Chapman}
{Chapman, S., } 1951, 
\textit{Arc. Metereol. Geophyis. Res.},  4, 369-390

\bibitem[Huang (2005)]{Huang}
{Huang, C.-M.\& Richmond A.D.} 1995,
\textit{J. Geophys. Res.}, 110, A05312 

\bibitem[Gjerloev (2012)]{Gjerloev}
{Gjerloev, J. W.} 2012, 
\textit{J. Geophys. Res.}, 117, A09213

\bibitem[Lopez (2014)]{Lopez}
{L\'opez, E, Maeda, G, et. al.} 2014, 
\textit{Sun and Geosphere}, 9, 31-34

\bibitem[Rigoti(1999)]{Rigoti}
{Rigoti, A., Chamalaun, N.B. \& Padilha, A. L.,} 1999, 
\textit{Earth Planets Space}, 51,115-128 

\bibitem[Takeda, 2002]{Takeda}
{Takeda, M., } 2002, 
\textit{J. Atmos. Sol. }, Terr. Phys. 64, 1617-1621

\bibitem[Yamazaki(2017)]{Yamazaki}
{Yamazaki, Y., Maute, A.} 2017,
\textit{Space Sci. Rev.}, 2017, 206-299
\end{thebibliography}
\end{document}